%% file: paper.tex
\documentclass[sigconf,authorversion=True]{acmart}

\AtBeginDocument{%
  }

\acmYear{2023}
\copyrightyear{2023}
\setcopyright{acmlicensed}
\acmConference[DEBS '23]{The 17th ACM International Conference on Distributed and Event-based Systems}{June 27--30, 2023}{Neuchatel, Switzerland}
\acmBooktitle{The 17th ACM International Conference on Distributed and Event-based Systems (DEBS '23), June 27--30, 2023, Neuchatel, Switzerland}
\acmPrice{15.00}
\acmDOI{10.1145/3583678.3596889}
\acmISBN{979-8-4007-0122-1/23/06}

\input{packages.tex}

\begin{document}

\title{No One Size (PPM) Fits All: \\ Towards Privacy in Stream Processing Systems}

\author{Mikhail Fomichev}
\authornote{Both authors contributed equally to this work.}
\affiliation{%
  \institution{TU Darmstadt}
  \country{}
}
\email{mfomichev@seemoo.tu-darmstadt.de}

\author{Manisha Luthra}
\authornotemark[1]
\affiliation{%
  \institution{TU Darmstadt \& DFKI}
  \country{}
}
\email{manisha.luthra@dfki.de}

\author{Maik Benndorf}
\affiliation{%
  \institution{University of Oslo}
  \country{}
}
\email{maikb@ifi.uio.no}

\author{Pratyush Agnihotri}
\affiliation{%
  \institution{TU Darmstadt}
  \country{}
}
\email{pratyush.agnihotri@kom.tu-darmstadt.de}

\renewcommand{\shortauthors}{Fomichev et al.}

\begin{abstract}
  \input{section/abstract.tex}
\end{abstract}



%

\maketitle

\input{section/introduction}
\input{section/related_work}
\input{section/vision}
\input{section/results}
\input{section/conclusion.tex}
\input{section/acknowledgement.tex}

\vspace{-2ex}
\bibliographystyle{ACM-Reference-Format}
\balance
\bibliography{bibliography}


\end{document}

%% file: packages.tex
\usepackage{booktabs} 
\usepackage[acronym, nowarn]{glossaries}
\makeglossaries

\usepackage{makecell}
\usepackage{multirow}

\usepackage{diagbox}
\usepackage[font=footnotesize,labelfont=bf]{caption}
\usepackage{xcolor,colortbl}

\usepackage{xspace}

\usepackage{todonotes}

\usepackage{tikz}

\newcommand{\rectangle}{\fboxsep0pt\fbox{\rule{1.4em}{0pt}\rule{0pt}{1.4ex}}}
\usepackage{halloweenmath} 
\usepackage{cleveref}

\usepackage{transparent}

\usepackage{adjustbox}

\newcommand{\tikzcircle}[2][red,fill=red]{\tikz[baseline=-0.5ex]\draw[#1,radius=#2] (0,0) circle ;}%
\usepackage{listings, xcolor}
\lstnewenvironment{codenv}{\lstset{language=SQL, 
basicstyle=\footnotesize}}{}
\usepackage{xcolor}
\definecolor{yelcir}{RGB}{255,192,0}
\definecolor{yelrim}{RGB}{188,140,0}
\definecolor{grecir}{RGB}{112,173,71}
\definecolor{grerim}{RGB}{80,126,50}
\definecolor{blucir}{RGB}{91,155,213}
\definecolor{blurim}{RGB}{65,113,156}
\definecolor{salcir}{RGB}{245,176,147}
\definecolor{salrim}{RGB}{237,125,49}

\definecolor{gr}{RGB}{198,196,215} 
\definecolor{pp}{RGB}{255,121,108} 
\definecolor{attred}{RGB}{192,0,0} 

\newcommand{\name}{\textsc{\textrm{Prinseps}}\xspace}

\input{abbreviations.tex}

%% file: abbreviations.tex
\newacronym{cep}{CEP}{Complex Event Processing}
\newacronym{put}{PUT}{privacy-utility tradeoff}
\newacronym{imu}{IMU}{inertial measurement unit}
\newacronym{iot}{IoT}{Internet of Things}
\newacronym{har}{HAR}{human activity recognition}
\newacronym{adl}{ADL}{activities of daily living}
\newacronym{ppm}{PPM}{privacy-preserving mechanism}
\newacronym{ml}{ML}{machine learning}
\newacronym{spe}{SPE}{stream processing engine}
\newacronym{qos}{QoS}{quality of service}
\newacronym{polp}{PoLP}{principle of least privilege}
\newacronym{dp}{DP}{differential privacy}
\newacronym{sps}{SPS}{stream processing system}
\newacronym{sp}{SP}{Stream Processing}
\newacronym{vae}{VAE}{variational autoencoder}
\newacronym{tee}{TEE}{trusted execution environment}
\newacronym{nlp}{NLP}{natural language processing}
\newacronym{ac}{AC}{access control}

\newacronym{hbc}{HBC}{honest-but-curious}

%% file: section/abstract.tex
\Glspl{sps} have been designed to process data streams in real-time, allowing organizations to analyze and act upon data on-the-fly, as it is generated. 
However, handling sensitive or personal data in these multilayered \glspl{sps} that distribute resources across sensor, fog, and cloud layers raises privacy concerns, as the data may be subject to unauthorized access  
and attacks that can violate user privacy, hence facing regulations such as the GDPR across the \gls{sps} layers. 
To address these issues, different \glspl{ppm} are proposed to protect user privacy in \glspl{sps}. 
Yet, selecting and applying such \glspl{ppm} in \glspl{sps} is challenging, since they must operate in real-time while tolerating little overhead. 
The multilayered nature of \glspl{sps} complicates privacy protection because each layer may confront different privacy threats, which must be addressed by specific \glspl{ppm}. 
To overcome these challenges, we present \name, our comprehensive privacy vision for \glspl{sps}. 
Towards this vision, we (1) identify critical privacy threats on different layers of the multilayered \gls{sps}, (2) evaluate the effectiveness of existing \glspl{ppm} in addressing such threats, and (3) integrate privacy considerations into the decision-making processes of \glspl{sps}. 

%% file: section/introduction.tex

\glsresetall
\vspace{-2ex}
\section{Introduction}
\label{sec:intro}
\textbf{Motivation.} \textit{Can we create a world without a disease?} 
This question was posed by Max Welling---a renowned \gls{ml} scientist in his visionary talk.\footnote{TEDx talk: \url{https://youtu.be/g9HA8A8tEUs} [Accessed on 26.05.2023].}
He underlined two requirements to achieve this goal: (1) privacy-preserving systems analyzing human-centric sensor data and (2) foundations in deep learning that enable better diagnosis and treatment of diseases. 
This work makes a step to fulfill the first requirement towards this noble goal using \glspl{sps}. 
\Glspl{sps} can serve as a core technology for analyzing sensor data collected by \gls{iot} devices in real-time. 
Recently, \glspl{sps} have adopted a \textit{multilayered approach} by distributing processing resources between \textit{sensor}, \textit{fog}, and \textit{cloud} layers to achieve lower processing latency and better utilization of resources~\cite{zeuch2020nebulastream}. 
Hence, we focus on multilayered \glspl{sps} in this work. 

The data streams that originate by sensing users or their environment inherently contain sensitive information, e.g., user's lifestyle, requiring \glspl{sps} that process such data to comply with privacy regulations, like the GDPR~\cite{stach2020bringing}. 
As crucially, users start to increasingly demand stricter control over the data collected by \gls{iot} applications about them~\cite{jin2022exploring}.
These points make privacy protection critical for the success of \glspl{sps} in the future. 

\begin{figure}
\centering
  \includegraphics[width=0.645\linewidth]{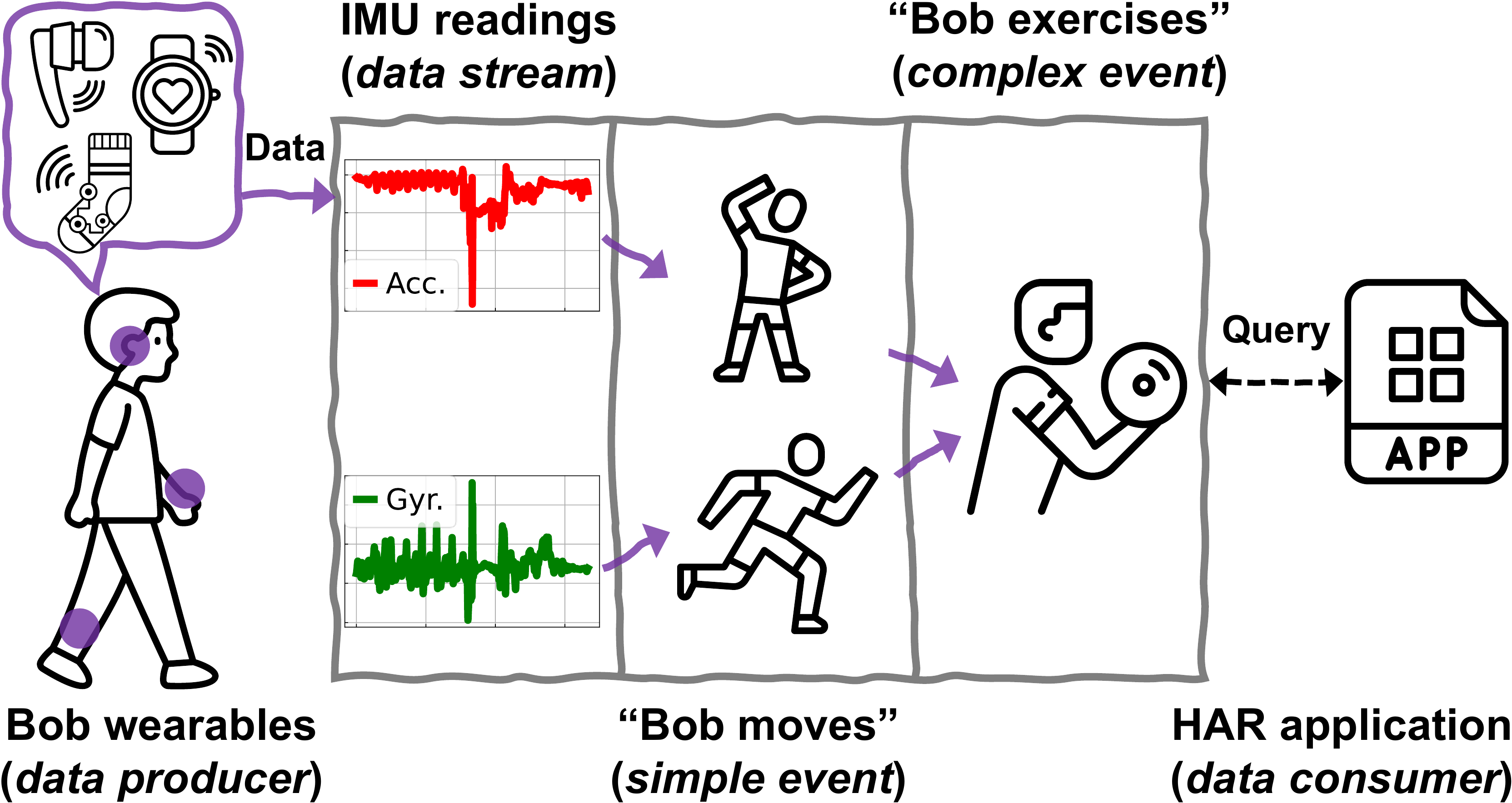}
  \vspace{-2ex}
  \caption{Typical workflow of an \gls{sps} based on our running example of a human activity recognition (HAR) application enabled by wearable sensors.}
  \vspace{-3.5ex}
  \label{fig:run-ex}
\end{figure}

\noindent
\textbf{Running Example.} To concertize our vision for privacy in multilayered \glspl{sps}, we pick a \gls{har} application enabled by wearable devices as our running  example in this work (cf.~\autoref{fig:run-ex}).
Specifically, a user: Bob carries wearable devices (e.g., smartwatch), known as \textit{data producers}, which collect \gls{imu} data, i.e., accelerometer and gyroscope readings.
Utilizing these data, the \gls{har} application, called the \textit{data consumer}, can monitor Bob's activity, like  exercising, for medical and health insurance purposes. 
This happens as follows: the stream of \gls{imu} data is input to an \gls{sps} which first classifies these data into \textit{simple events} (e.g., ``Bob moves'') and then combines such simple events to detect \textit{complex events}, like ``Bob exercises''. 
The \gls{har} application is mainly interested in monitoring these complex events, hence it defines a \textit{continuous query} to identify them in the ceaseless stream of the \gls{imu} data. 

Our running example exhibits that the \gls{sps} scrutinizes Bob's data on different granularity: from raw \gls{imu} data to complex events. 
This clearly has privacy concerns, like the \gls{har} application can detect not only \textit{public} complex events (or patterns) that are needed for its functionality, e.g., ``Bob exercises'', but also \textit{private} complex events (or \textit{patterns}), like ``taking medicine'' by Bob, which is revealed as a sequence of ``swallow'' $\rightarrow$ ``drink'' $\rightarrow$ ``lay down'' events.

\noindent
\textbf{Contributions.} The topic of privacy protection in \glspl{sps} has recently emerged, yet the privacy issues stemming from \gls{sps}'s multilayeredness remain underexplored~\cite{plagemann2022towards, stach2020bringing}.
As such, the \glspl{ppm} utilized in \glspl{sps} are typically adopted from the database community, requiring a major redesign to satisfy the real-time nature of \glspl{sps}~\cite{palanisamy2018preserving}, while several \glspl{ppm} developed specifically for \glspl{sps} are still not mature, which makes it difficult to holistically apply them in \glspl{sps}~\cite{stach2020bringing}. 
We identify \textit{three challenges} of using existing \glspl{ppm} in multilayered \glspl{sps}: (1) threats to user privacy are not systematically understood in these \glspl{sps}, complicating the selection of appropriate \glspl{ppm}; (2) \glspl{ppm} must \textit{not only} be applied holistically, i.e., at each layer of the multilayered \gls{sps} to provide strong privacy protection, \textit{but also} satisfy real-time needs of \glspl{sps}, imposing little overhead; (3) the \glspl{ppm}, selected and applied in this fashion, should enable a \textit{\gls{put}}, which is tunable depending on \gls{sps}'s user privacy needs and application \textit{\gls{qos}} requirements.
These challenges unfold \textit{two research questions} that we seek to answer in this work: \\
\textbf{RQ1:} What are the most critical threats to user privacy in multilayered \glspl{sps} used in \gls{iot} applications? \\
\textbf{RQ2:} How to systematically select, customize, and apply \glspl{ppm} that address these threats while enabling a \gls{put}?

Addressing the above questions and inspired by~\cite{plagemann2022towards}, we present our vision for holistic privacy protection in multilayered \glspl{sps} called \name---\textit{\underline{Pr}ivacy \underline{in} \underline{S}tr\underline{e}am \underline{P}rocessing \underline{S}ystems}.
To pursue this vision, we (1) review the state of privacy protection in \glspl{sps} (\autoref{sec:rwork}); (2) identify critical privacy threats and their conduct in multilayered \glspl{sps}, which allows us to shape \name' design (\autoref{sec:vision}); and (3) provide the preliminary evaluation of \name along with further research opportunities (\autoref{sec:results}). 
Despite focusing on the \gls{iot} use cases for \glspl{sps} to concretize our vision, we perform the above three steps in a generic manner, making \name generalizable to other \glspl{sps}' applications.
Our results show that there is \textit{no} one-size-fits-all solution for privacy in \glspl{sps}, i.e., different \glspl{ppm} have their up- and downsides (which we investigate), while being applicable in various parts of the multilayered \gls{sps}.

%% file: section/related_work.tex

\vspace{-2ex}
\section{Related Work}
\label{sec:rwork}
\noindent
\textbf{Privacy in Stream Processing.} The subject of privacy protection in \glspl{sps} has started receiving attention from the research community, yet existing works mainly focus on preserving privacy in \textit{one} of the following directions within \glspl{sps}: (1) raw data streams or (2) private patterns, whereas a holistic view on privacy is still missing.
For \textit{raw data streams}, prior research \textit{adopted} \glspl{ppm} from the database community, like \gls{dp}, k-anonymity, and l-diversity~\cite{cao2010castle, guo2013fast, chen2017pegasus}. 
Such \glspl{ppm} seek to protect single events (e.g., ``Bob moves'' from \autoref{fig:run-ex}) by grouping them into clusters---within which these events become indistinguishable for the data consumer of an \gls{sps}, like a \gls{har} application. 
There exist two issues with such adopted \glspl{ppm}: (1) they introduce extra overhead due to clustering, harming the real-time operation of \glspl{sps}~\cite{fioretto2019optstream}; (2) most works assume that sensitive events are revealed via simple \textit{count} queries~\cite{cao2010castle, quoc2017privapprox, chen2017pegasus}, which does not reflect real-world \glspl{sps}---where both more complex \textit{queries on} raw data streams and \textit{access to} such streams can become 

\noindent
attack vectors to violate user privacy~\cite{johnson2020chorus, dwarakanath2017trustcep}.

There exist a few \glspl{ppm} protecting \textit{private patterns}, e.g., ``taking medicine'' by Bob.
\cite{palanisamy2018preserving} presents a \gls{ppm} leveraging event reordering to conceal private patterns.
In~\cite{dwarakanath2017trustcep}, a \gls{ppm} that enforces \gls{ac} on private patterns is proposed. 
This \gls{ppm} estimates trust among nodes within an \gls{sps}, allowing only trusted nodes to access private patterns.
While being important preliminary research, the \glspl{ppm}~\cite{palanisamy2018preserving} and~\cite{dwarakanath2017trustcep} lack robust privacy guarantees and rely on strong assumptions, like the availability of trust information in an \gls{sps}. 

A related line of work to privacy in raw data streams and private patterns is \textit{privacy-preserving stream analytics}, which enforces \gls{ac} on \gls{sps}'s queries by means of cryptography~\cite{burkhalter2020timecrypt}.  
While these \glspl{ppm} protect against malicious attackers, they become less relevant under our threat model (cf.~\autoref{sec:vision}), where such \glspl{ppm} using encryption and authentication are assumed to be in place. 
Similar to the stream analytics, \textit{privacy preservation in publish/subscribe systems} heavily relies on \gls{ac}-based \glspl{ppm} to shield sensitive data flows~\cite{onica2016confidentiality}. 
Another trend in privacy protection that benefits all of the described \glspl{ppm} is leveraging \glspl{tee} to safeguard the runtime of \glspl{ppm} within an \gls{sps} against malicious components~\cite{scopelliti2023end}.

\noindent
\textbf{Privacy-utility Tradeoff.} 
The above \glspl{ppm} only loosely consider a \gls{put} when applied in \glspl{sps}, i.e., the impact of a \gls{ppm} is, in the \textit{best case}, evaluated as an interplay between one privacy and one utility metric in a specific scenario. 
Yet, the tradeoff between these metrics, its fine-tuning, and generalizability to other scenarios receive little attention. 
A few recent works present first systematic approaches to balance a \gls{put}, which can either be achieved via an optimization problem~\cite{biswas2022three} or through user input~\cite{erdemir2022active}.
However, the applicability of such approaches has \textit{not yet} been explored in the context of \glspl{sps}.

\noindent
\textbf{Research Gap.} As shown above, the privacy protection in \glspl{sps} has been applied in an \textit{ad-hoc fashion}, adopting existing or devising new \glspl{ppm} to address isolated privacy issues.
This happens due to limited understanding of critical privacy threats and their occurrence inside \glspl{sps}.
Thus, it is difficult to justify the selection, customization, and application of \glspl{ppm} addressing those threats while preserving the utility of \glspl{sps}. 
Without such justification, it remains unclear \textit{where and how} to apply even the golden standard \glspl{ppm}, like \gls{dp}~\cite{ren2022ldpids}. 

%% file: section/vision.tex

\vspace{-2ex}
\section{\name: Vision for Privacy in \glspl{sps}}
\label{sec:vision}
To approach the vision of \name, we take four steps: (1) identify critical privacy threats and their conduct in a multilayered \gls{sps}; (2) use this knowledge to select exemplary \glspl{ppm} that we evaluate to find their suitability for \name; (3) show how to customize these \glspl{ppm}, leveraging user privacy policies, to facilitate a \gls{put}; and (4) formulate the principles of applying \glspl{ppm} in \name to enable a holistic privacy protection in multilayered \glspl{sps}. 

We envision the following \textit{success criteria} for \name: (1) simultaneous protection against critical privacy threats in a multilayered \gls{sps}, allowing its users to control the protection granularity as well as fine-tune the \gls{put}; (2) seamless extendibility of \name with new \glspl{ppm} to address existing and/or novel privacy threats.

\begin{figure}
\centering
  \includegraphics[width=0.509\linewidth]{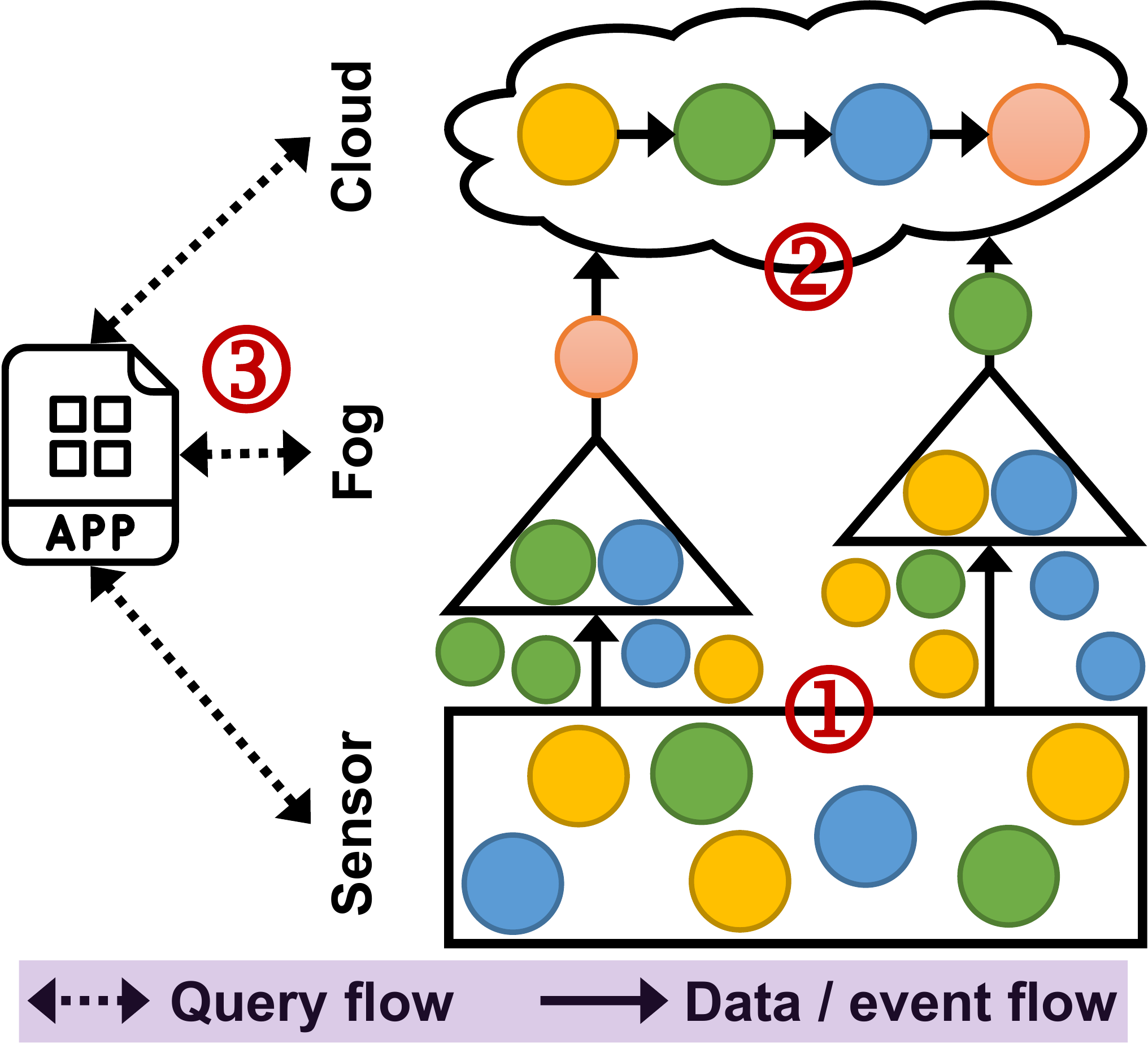}
  \vspace{-2ex}
  \caption{Multilayered \gls{sps} architecture used in \name, consisting of nodes placed on sensor ($\rectangle$), fog ({\Large$\bigtriangleup$}), and cloud ($\mathcloud$) layers;  \tikzcircle[blurim,fill=blucir]{4.0pt},  \tikzcircle[yelrim,fill=yelcir]{4.0pt}, \tikzcircle[grerim,fill=grecir]{4.0pt} denote data or simple events from different sensors, while \tikzcircle[salrim,fill=salcir]{4.0pt} is an inferred complex event. \textcolor{attred}{\raisebox{.5pt}{\textcircled{\raisebox{-.9pt} {1}}}} to \textcolor{attred}{\raisebox{.5pt}{\textcircled{\raisebox{-.9pt} {3}}}} mark the occurrence of critical privacy threats in this architecture.}
  \label{fig:priv-arch}
  \vspace{-3.5ex}
\end{figure}

\noindent
\textbf{Privacy Threats.} 
To identify critical privacy threats and how they occur within a multilayered \gls{sps}, whose architecture is shown in~\autoref{fig:priv-arch}, we use our running example. 
In~\autoref{fig:priv-arch}, the \gls{imu} \textit{data flows} from Bob's wearables, e.g., a smartwatch (\textit{sensor layer}), up into fog nodes, like a smart hub, where these data are classified into simple events, e.g., ``Bob moves'' (\textit{fog layer}), that are pushed up to the cloud, where complex events, like ``Bob exercises'', are ultimately inferred from such aggregated simple events (\textit{cloud layer}).
To monitor Bob's activity, the \gls{har} application queries the \gls{sps}---we assume the \textit{query flow} can happen between the \gls{har} application and \textit{any} of the \gls{sps}'s layers: sensor, fog, and cloud (cf. dotted arrow lines in~\autoref{fig:priv-arch}); this gives flexibility to \glspl{sps}, e.g., in terms of query result granularity.

We use an \textit{\gls{hbc}} adversary model to identify critical privacy threats in multilayered \glspl{sps}, due to its realism: here, we assume that an \gls{sps} node on each layer (sensor, fog, cloud) and data consumer (\gls{har} application) can be \gls{hbc}~\cite{palanisamy2018preserving, quoc2017privapprox}.
Such nodes and data consumers conform to \gls{sps}'s functionality, i.e., they detect expected events faithfully and do not compromise the correctness of queries, but seek to learn extra information about Bob, like habits, invading his privacy. 
We differentiate \textit{attack vectors} between \gls{hbc} nodes and data consumers: the former scrutinize Bob's sensor data and events during their processing, while the latter craft queries to reveal Bob's sensitive events, e.g., ``taking medicine''. 

Based on our threat model, we conduct research utilizing related work analysis, self-expertise, discussions with privacy practitioners, and review of best privacy practices\footnote{These steps follow the current state of the art on privacy threat modeling~\cite{bloom2022privacy}.} to determine generic privacy threats in multilayered \glspl{sps}: (1) \textit{sensitive attributes}, (2) \textit{private patterns}, and (3) \textit{invasive queries}.
Note that the found three threats are not claimed to be exhaustive, and we presume the \glspl{ppm} preventing malicious attackers, like encryption (cf.~\autoref{sec:rwork}), to be deployed---such \glspl{ppm} alone cannot thwart \gls{hbc} adversaries which could, e.g., legitimately access encryption keys.
In the following, we detail the three privacy threats identified in multilayered \glspl{sps}, as they help us to shape the design of \name.

\textit{(1) Sensitive Attributes.} 
The sensitive attributes, like Bob's weight, can be inferred from a few seconds of \gls{imu} data using deep learning~\cite{hajihassnai2021obscurenet}. 
Such amount of data is available on sensors and fog devices (e.g., smartwatch and smart hub), whose hardware is performant to carry inferences of sensitive attributes in real-time~\cite{hajihassnai2021obscurenet}. 
Hence, this threat happens on \gls{hbc} sensor- and fog-layer nodes, during processing of Bob's \gls{imu} data (cf.~\raisebox{.5pt}{\textcircled{\raisebox{-.9pt} {1}}} in~\autoref{fig:priv-arch}). When choosing the \glspl{ppm} to tackle the inference of sensitive attributes, we try to achieve \textit{low latency} as our primary utility metric of an \gls{sps}, since low latency is the main advantage of processing data near its producers~\cite{zeuch2020nebulastream}. 

\textit{(2) Private Patterns.} The private patterns, such as ``taking medicine'' by Bob, are complex events detectable by an \gls{sps} via sequence matching of (simple) events comprising this pattern. 
Hence, this threat materializes on fog and cloud layers, where the chronological sequences of events are available (cf.~\raisebox{.5pt}{\textcircled{\raisebox{-.9pt} {2}}} in~\autoref{fig:priv-arch}). 
We concur with prior research that issuing \gls{hbc} queries to fog and cloud nodes is the most feasible strategy for revealing private patterns~\cite{palanisamy2018preserving}. 
When selecting the \glspl{ppm} to conceal private patterns, we seek \textit{high accuracy} of public complex events (e.g., ``Bob exercises''), that should be detectable by an \gls{sps}, as our main utility metric. 

\textit{(3) Invasive Queries.} The invasive queries aim to reveal sensitive information about Bob, like lifestyle, going beyond private patterns. 
Such threats happen during query processing by \gls{hbc} nodes which can, e.g., augment a query on Bob's exercising with the simple filter condition: \texttt{time > 9am} \texttt{AND} \texttt{time < 6pm} to disclose the fact of Bob's exercising during working hours to his employer.
Because \name allows queries to be processed by sensor, fog, and cloud layers, this threat can occur on any of them (cf.~{\textcircled{\raisebox{-.9pt} {3}}} in~\autoref{fig:priv-arch}). 
When choosing the \glspl{ppm} to address invasive queries, we consider \textit{both} low latency and high accuracy of public (complex) events to be our target utility metrics of an \gls{sps}.

\noindent
\textbf{Exemplary \glspl{ppm} for \name.} After describing critical privacy threats and their conduct in multilayered \glspl{sps}, we utilize (1) assets that need protection, (2) target utility metrics, and (3) 
the computing capabilities of \gls{sps}'s layers as our criteria to select exemplary \glspl{ppm} for \name.
We identify \textit{three types of \glspl{ppm}} that satisfy our criteria while being frequently used by real-world applications~\cite{erdemir2022active, hajihassnai2021obscurenet}: machine learning (\gls{ml})-, differential privacy (\gls{dp})-, and access control (\gls{ac})-based \glspl{ppm}.
We showcase how these \glspl{ppm} address sensitive attributes, private patterns, and invasive queries threats in multilayered \glspl{sps} (cf.~\autoref{sec:results}), shedding light on which other information can be utilized by \name to select \glspl{ppm}.  

\noindent
\textbf{Privacy Policies: Customizing \glspl{ppm} to Enable the \gls{put}.} There exists clear evidence that \glspl{ppm} can improve their \gls{put} by protecting the privacy of ``what users care for''~\cite{erdemir2022active, tex2022swellfish}. 
Hence, identifying \textit{user privacy needs} can help \name to customize \glspl{ppm}, making their parameterization and application \textit{selective}.
This enhances utility, as protection levels of \glspl{ppm} (e.g., amount of noise added to data) and their usage frequency are limited to user privacy needs, which also allows fine-tuning the \gls{put}.

\begin{table}
\scriptsize
 \caption{Requirements for formulating privacy policies in \name.}
 \vspace{-2ex}
  \label{tab:priv-req}
  \begin{tabular}{|l|l|}
    \toprule
    Requirement & Explanation \\
    \midrule
    \textbf{R1}: Transparency & Users understand purposes for which their data are utilized. \\
    \textbf{R2}: Accessibility & Users configure their privacy in an easy and intuitive manner. \\
    \textbf{R3}: Proactiveness & Users are assisted in configuring their privacy. \\
    \textbf{R4}: Awareness & Users are made aware of privacy threats they did not consider. \\
    \textbf{R5}: Granularity & Users control their privacy in a fine-grained manner. \\
   \bottomrule
\end{tabular}
\vspace{-3.5ex}
\end{table}

\name captures users' privacy needs in the form of a privacy policy. 
The literature shows various requirements for a user-defined privacy policy~\cite{stach2020bringing, jin2022exploring}---we extract the most relevant for \name in~\autoref{tab:priv-req}.
We create the privacy policy based on \textit{static} and \textit{dynamic rules}~\cite{stach2020bringing}, as explained using our running example. 
Respecting \textbf{R1:} \textit{Transparency}, a \gls{har} application must brief Bob that \gls{imu} data from his wearables is collected to monitor his exercise. 
Bob can also be prompted to provide his weight for a better \gls{qos}---if he decides \textit{not} to disclose it, then such a privacy need must be intuitively expressive, satisfying \textbf{R2:} \textit{Accessibility}. 

\name utilizes \textit{policy rules} to map Bob's privacy needs, like a static policy rule to conceal his weight. 
Such policy rules can be predefined in \name by domain privacy experts, e.g., in the form of trigger-action statements. 
\name finalizes Bob's privacy policy by clarifying and making suggestions to select best-matching policy rules, as per \textbf{R3:} \textit{Proactiveness}. 
Despite hiding his weight, Bob must be alarmed that \textit{its inference} is still feasible via \gls{imu} data, fulfilling \textbf{R4:} \textit{Awareness}. 
Hence, \name offers Bob to either apply a \gls{ppm}(-s) against this threat or to accept it. 
In the first case, Bob gets informed about \gls{ppm}'s impact on utility, and he is guided towards tuning his \gls{put}, following \textbf{R5:}~\textit{Granularity}.
For example: the used \gls{ppm} can, by default, reduce weight inference to 10\% at the cost of recognizing 90\% of Bob's activity correctly; 
Bob can adjust this \gls{put} to, e.g., 1\% weight inference vs. 80\% of activity recognition accuracy. 

Bob may need to change the usage of a specific \gls{ppm}, like its \gls{put}, based on his \textit{context}, e.g., time and location.
\name achieves this via dynamic rules that override Bob's static policy rules, relying on his feedback in a situation, e.g., Bob allows the \gls{har} application to access his weight \textit{only if} he is at a doctor's office. 

\noindent
\textbf{Principles of Applying \glspl{ppm}.} To enable holistic privacy protection in \glspl{sps}, the application of \glspl{ppm} in a \textit{privacy-by-design} manner is as vital as \glspl{ppm}' privacy guarantees~\cite{stach2020bringing}. 
\name follows three such privacy-by-design principles: \textbf{P1:} the \textit{\Gls{polp}}, \textbf{P2:} \textit{Applying a \gls{ppm} ``early on''}, and \textbf{P3:} \textit{Complementarity of \glspl{ppm}}. 
With \textbf{P1}, data minimization is enforced, like forbidding the smart home apps to access raw sensor data, since they only require high-level events for their functionality~\cite{jin2022exploring}. 
Using \textbf{P2}, we demand that a \gls{ppm} is applied \textit{as close} towards the source of the threat \textit{as possible}, thus it will not propagate (e.g., to the next layer in an \gls{sps}) or escalate. 
\textbf{P3} stipulates that each known threat must, by default, be treated by a \gls{ppm}.
Yet, complex threats may not be preventable by a single \gls{ppm}: in this case, the combination of \glspl{ppm} that reinforce each other should be applied to tackle the complex threats. 
To \textit{avoid circumventing} \glspl{ppm} used in \name via misconfiguration, we, on top of \textbf{P1}--\textbf{P3}, demand such \glspl{ppm} to be \textit{deployed as per best security practices}, like running \glspl{ppm} inside \glspl{tee} and obfuscating them~\cite{bloom2022privacy}.

%% file: section/results.tex

\vspace{-2ex}
\section{Shaping \name: Results}
\label{sec:results}

\begin{figure}
    \centering
    \includegraphics[width=0.839\linewidth]{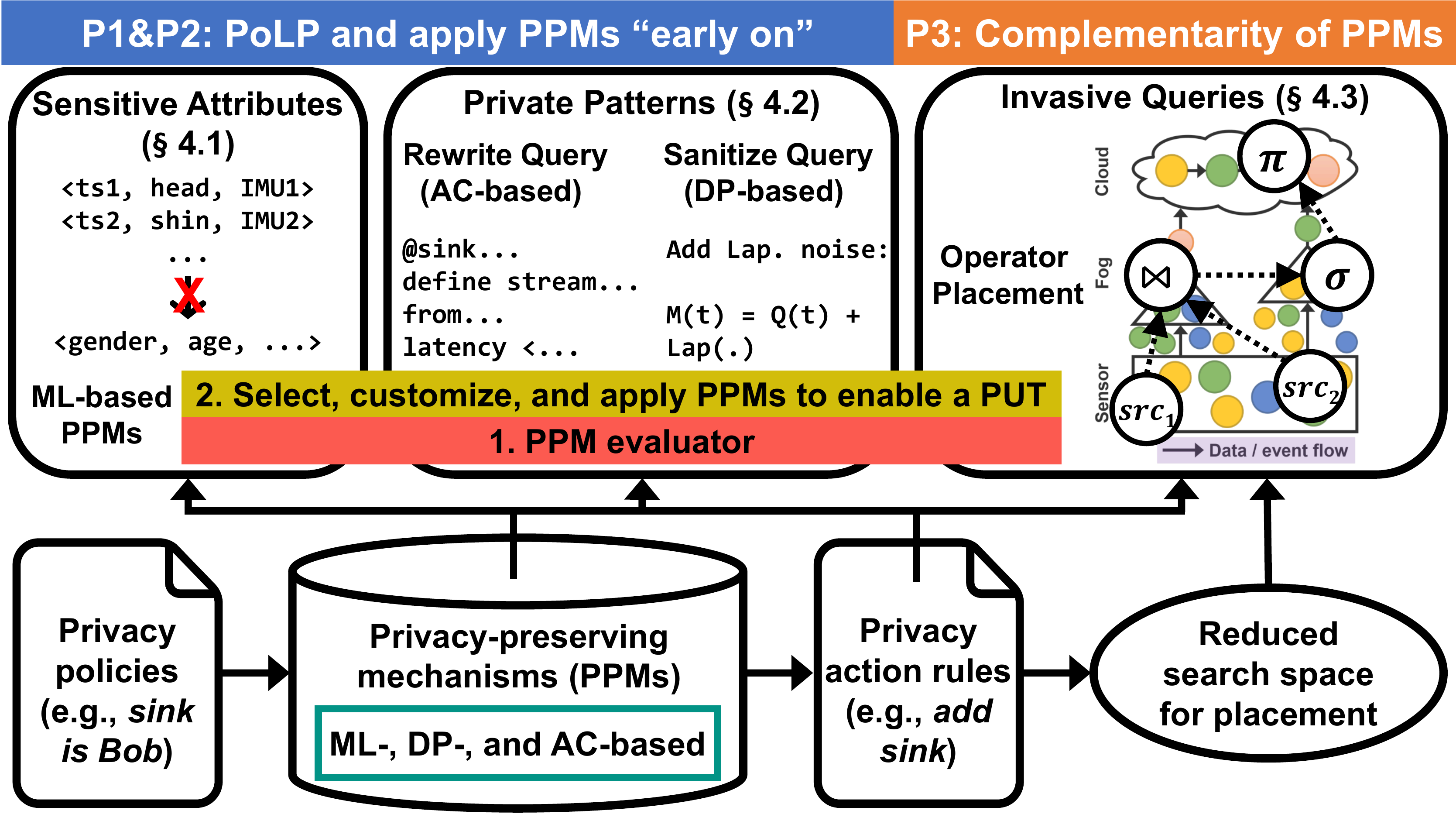}
    \vspace{-2ex}
    \caption{Overview of \name privacy-preserving architecture for \glspl{sps}.}
    \label{fig:overview}
    \vspace{-3.5ex}
\end{figure}

\autoref{fig:overview} shows an overview of \name whose goal is to \textit{holistically} address critical privacy threats in multilayered \glspl{sps} while enabling a \gls{put}. 
\name has the following workflow: the \textit{pool of (exemplary) \glspl{ppm}} tackling these critical threats is maintained, i.e., novel \glspl{ppm} are included as new privacy threats appear, and outdated \glspl{ppm} get removed. 
From this pool, the \name' component called the \textit{\gls{ppm} evaluator} picks a candidate \gls{ppm} to address a specific privacy threat based on \textit{criteria} such as: privacy guarantees, runtime performance, utility metrics, resource requirements, scalability, and ease of setup; this list is nonexhaustive.
Then, \name customizes the candidate \gls{ppm} leveraging a \textit{user privacy policy} to fulfill his/her privacy needs while minimizing the impact on utility by restricting \gls{ppm}'s usage to such privacy needs and its level of protection---to user's preferences.
Finally, the customized \gls{ppm} is deployed at a specific \gls{sps} layer, i.e., sensor, fog, or cloud (cf.~\autoref{fig:priv-arch}).

The workflow of \name is \textit{dynamic}, i.e., each component can receive updates and be rerun, like including a new \gls{ppm} to the pool can trigger the evaluator to reassess this \gls{ppm} as a better candidate to tackle a privacy threat, or adding a fresh privacy policy rule will require \name to recustomize the \gls{ppm}(-s) already in use. 

In the following, we report our findings on how \name utilizes state-of-the-art \gls{ml}-, \gls{dp}-, and \gls{ac}-based \glspl{ppm} to address sensitive attributes, private patterns, and invasive queries threats (cf.~\autoref{sec:vision}). 
We also discuss open challenges and research opportunities to shape \name and, more broadly, the privacy landscape in \glspl{sps}.
Sections~\ref{subsec:sen-attr} through \ref{subsec:inv-quer} cover each of the above privacy threats and are \textit{structured as follows}: we (1) justify the selection of most suitable \gls{ppm}(-s) against the threat, using our principles of applying \glspl{ppm} listed in~\autoref{sec:vision}, findings from related work, and self-expertise; (2) report evaluation results for each \gls{ppm} tackling the corresponding privacy threat; and (3) present opportunities for future research. 

\vspace{-2ex}
\subsection{Conceal Sensitive Attributes} 
\label{subsec:sen-attr}
\noindent
\textbf{\gls{ppm} Selection.} Among \gls{ml}-, \gls{dp}-, and \gls{ac}-based \glspl{ppm}, we see that the former fits best to conceal sensitive attributes, as they: (1) work directly on raw data streams (\textbf{P2}), minimizing the information on the sensitive attribute, like weight (\textbf{P1}); and (2) show near real-time performance~\cite{hajihassnai2021obscurenet}, allowing \name to meet the utility goal related to this threat---low latency. 
Contrarily, \gls{dp}-based \glspl{ppm} target raw data streams by adding noise to query results on such data (cf.~\autoref{sec:rwork}), violating \textbf{P2}, whereas injecting noise directly into raw data 
either degrades its accuracy (and thus utility) or leads to significant noise scalability issues~\cite{fioretto2019optstream}. 
The \gls{ac}-based \glspl{ppm} could, in principle, conceal sensitive attributes by limiting access to parts of raw data, satisfying \textbf{P1} and \textbf{P2}.
Yet, sensitive attributes are embedded in raw data (e.g., gender in the \gls{imu} data of Bob)~\cite{hajihassnai2021obscurenet}, making it difficult to formulate such \gls{ac}-rules.   

To see how \name can utilize \gls{ml}-based \glspl{ppm}, we review one recent \gls{ppm}---\textit{ObscureNet} that uses the \textit{\gls{vae}} neural network to transform \gls{imu} data, hiding sensitive attributes in near real-time~\cite{hajihassnai2021obscurenet}. 
This \gls{ppm} reduces the success of inferring a sensitive attribute, like gender, from 90\% to 50\% while keeping the accuracy of public \gls{har} events, searched by an \gls{sps}, above 90\%.

\noindent
\textbf{Results: \gls{ml}-based \glspl{ppm}.} We seek to understand the (1) \gls{put} and (2) runtime performance of such \glspl{ppm} in the real-world \gls{sps} case of several wearable devices, as per our running example.
Therefore, we evaluate ObscureNet on the \textit{RealWorld (\gls{har})} dataset that contains \gls{imu} data from \textit{five devices} located as such: head, upper arm, waist, thigh, and shin~\cite{sztyler2016body}; originally, ObscureNet is validated on the \gls{imu} data coming from a single thigh-worn device~\cite{hajihassnai2021obscurenet}. 

For comparability, we use ObscureNet\footnote{\url{https://github.com/sustainable-computing/ObscureNet} [Accessed on 26.05.2023].} with the same parameters as in~\cite{hajihassnai2021obscurenet}; the sampling rate of \gls{imu} data from RealWorld (\gls{har}) is 50 Hz. 
\autoref{tab:gen-inf} shows the \textit{gender inference} accuracy on the five body locations (i.e., head to shin) for four \gls{har} events: walking, standing, jogging, and climbing upstairs, which must be detected by the \gls{sps}.
We see that gender inference drops from over 90\% to below 20\% after applying ObscureNet, being in line with~\cite{hajihassnai2021obscurenet}. 
This behavior is consistent across different body locations and \gls{har} events; our results for \textit{weight inference} are similar.
Nevertheless, we reveal that ObscureNet can lower the accuracy of \gls{har} events by up to 50\% (cf.~ \autoref{tab:act-inf}), harming the \gls{sps}'s utility, which was unseen in~\cite{hajihassnai2021obscurenet}.

We assess the runtime of ObscureNet for several devices reusing the benchmark of~\cite{hajihassnai2021obscurenet} carried on off-the-shelf hardware: Raspberry Pi 3. 
From this benchmark, we calculate the time budget for running ObscureNet on our \gls{imu} data to be 200 ms, while ObscureNet needs around 100 ms to conceal sensitive attributes on these data. 
Hence, for the \gls{imu} data of 50 Hz, ObscureNet can protect privacy of data streams from \textit{two devices} simultaneously in near real-time.   

\begin{table}
\scriptsize
\centering
\caption{Gender inference accuracy on five body locations for four \gls{har} events before and after applying ObscureNet.}
\vspace{-2ex}
        \label{tab:gen-inf}
  \begin{tabular}{|c|*{4}{c|}}
	\toprule
	\diagbox{Location}{Accuracy}{Event}& \makecell{Walking \\ (b / a, $\%$)}  & \makecell{Standing \\ (b / a, $\%$)}  &    \makecell{Jogging \\ (b / a, $\%$)}  & \makecell{Upstairs \\ (b / a, $\%$)} \\
	\midrule
	Head & 99.1 / 14.7 & 92.3 / 16.1 & 98.9 / 17.3 & 97.9 / 17.4  \\
	Upper arm & 99.5 / 7.1 & 94.0 / 41.0 & \cellcolor{gray!35}{98.3 / \textbf{10.8}} & 99.0 / 22.5  \\
	Waist & 99.4 / 13.1 & \cellcolor{gray!35}{90.2 / \textbf{15.3}} & 93.8 / 20.6 & 98.1 / 35.3  \\
	Thigh & \cellcolor{gray!35}{95.0 / \textbf{6.2}} & 80.0 / 31.2 & 91.3 / 12.0 & \cellcolor{gray!35}{95.5 / \textbf{16.8}} \\
	Shin & 92.1 / 20.1 & 73.0 / 45.2 & 93.1 / 28.7 & 92.6 /  22.1 \\
	\hline
	Combined & 97.0 / 29.4 & 85.9 / 33.6 & 95.2 / 36.7 & 94.7 / 36.9 \\
	\bottomrule
\end{tabular}
  \centering
    \center{(b / a, $\%$) -- inference accuracy in $\%$ \textit{before} and \textit{after} applying ObscureNet.}
    \vspace{-3.5ex}
\end{table}

\begin{table}
\scriptsize
  \caption{Event inference accuracy on five body locations for four \gls{har} events before and after applying ObscureNet.}
  \vspace{-2ex}
      \label{tab:act-inf}
\centering
  \begin{tabular}{|c|*{4}{c|}}
	\toprule
	\diagbox{Location}{Accuracy}{Event}& \makecell{Walking \\ (b / a, $\%$)}  & \makecell{Standing \\ (b / a, $\%$)}  &    \makecell{Jogging \\ (b / a, $\%$)}  & \makecell{Upstairs \\ (b / a, $\%$)} \\
	\midrule
	Head & 91.1 / 39.0 & 90.0 / 66.4 & 98.4 / 83.4 & 98.0 / 89.5  \\
	Upper arm & 92.4 / 66.8 & 90.8 / 78.4 & 95.6 / 87.3 & 96.4 / 86.1  \\
	Waist & \cellcolor{gray!35}{97.7 / \textbf{93.7}} & 82.7 / 69.8 & 98.4 / 89.6 & 97.8 / 80.0 \\
	Thigh & 92.0 / 75.8 & 92.9 / 66.1 & 95.5 / 88.7 & 97.6 / 95.1  \\
	Shin & 92.6 / 60.5 & \cellcolor{gray!35}{94.3 / \textbf{92.8}} & 94.7 / 87.9 & \cellcolor{gray!35}{96.6 / \textbf{94.4}}  \\
	\hline
	Combined & 90.5 / 37.3 & 82.1 / 62.4 & \cellcolor{gray!35}{93.7 / \textbf{87.6}} & 93.5 / 83.5  \\
	\bottomrule
\end{tabular}
 \centering
    \center{(b / a, $\%$) -- inference accuracy in $\%$ \textit{before} and \textit{after} applying ObscureNet.}
    \vspace{-3.5ex}
\end{table}

\noindent
\textbf{Research Opportunity.} 
Our results indicate that \gls{ml}-based \glspl{ppm}, like ObscureNet, favor privacy over utility in realistic \gls{sps} cases with several wearables. 
Hence, utility---\textit{both} in terms of accuracy of \gls{har} events and runtime performance (i.e., for lower latency)---needs to be improved, which requires an \textit{in-depth study} of such recent \glspl{ppm} to adjust their \gls{put}. 
This twofold utility can be enhanced through \textit{federated learning}, leveraging data streams from different devices \cite{yang2022blinder}.
The runtime performance of ObscureNet boosts by lowering the data sampling rate, like by two times if we go from 50 to 25 Hz.
Thus, \gls{ml}-based \glspl{ppm} should be further explored towards utilizing task-aware knowledge. Studying \name' \textit{generalizability}, we spot a research avenue---to investigate sensitive attributes in nonhuman-centric environments (e.g., smart factories) and evaluate the efficacy of existing \gls{ml}-based \glspl{ppm} in concealing such sensitive attributes.


\vspace{-2ex}
\subsection{Protect Private Patterns} 
\label{subsec:prv-patt}
\noindent
\textbf{\gls{ppm} Selection.} From \gls{ml}-, \gls{dp}-, and \gls{ac}-based \glspl{ppm}, we find that the latter two are best suited to protect private patterns, like ``taking medicine'' in our running example.
The reasons are as follows: both \glspl{ppm} have been used on queries~\cite{tex2022swellfish, ibm2023query} while satisfying \textbf{P1} and \textbf{P2}. 
Considering \gls{dp}-based \glspl{ppm}, they minimize private information in the query result by \textit{adding noise} to it (sanitization) \textit{after} processing the query, but before returning its response~\cite{tex2022swellfish}.  
While \gls{ac}-based \glspl{ppm} \textit{rewrite queries before} processing them, restricting access to sensitive data leaked by the query result~\cite{ibm2023query}.
Thus, these \glspl{ppm} can also be complementary, as per \textbf{P3}. 
We have not found any \gls{ml}-based \glspl{ppm} that are applicable on queries to protect private patterns.

\noindent
\textbf{Results: \gls{dp}-based \glspl{ppm}.} We (1) determine the most suitable class of \gls{dp}-based \glspl{ppm} for protecting private patterns and (2) instantiate one such \gls{ppm}~\cite{tex2022swellfish}, exhibiting how it can be leveraged by \name. 

To date, there exist three classes of \gls{dp}-based \glspl{ppm} adding noise at the granularity of: events, windows of events, and users---denoted as \textit{event-level}, \textit{w-event level}, and \textit{user-level}  \glspl{ppm}~\cite{ren2022ldpids}.
The first class targets single events, e.g., ``Bob drinks'', and not their relationships, making these \glspl{ppm} unsuitable for our purpose. 
As for the user-level \gls{dp}, it protects the entire data stream of a person but requires many users to be involved, rendering such \glspl{ppm} infeasible in single-user 
cases (as our running example) due to a reduced \gls{put}~\cite{ren2022ldpids}. 
Instead, we study w-event level \glspl{ppm} that conceal a series of events within a time window, allowing a better \gls{put} for protecting private patterns.  

We instantiate a novel w-event level \gls{ppm}: \textit{Swellfish privacy}~\cite{tex2022swellfish}, for our running example to assess its impact on the \gls{put}. 
This \gls{ppm} improves on existing w-event level \gls{dp} methods by adding noise to queries \textit{selectively}, i.e., only within a \textit{relevance interval} provided by a user (e.g., Bob), and \textit{not} universally to every window of length $w$ events occurring in the data stream. 
The relevance interval specifies a timeframe inside which a private pattern can happen, like ``taking medicine'' by Bob around mealtimes.
Being selective in adding noise, allows~\cite{tex2022swellfish} to attain between \textit{one to three orders of magnitude} better \gls{put} than the state of the art on w-event level \gls{dp}.
Swellfish privacy is thus an exemplary \gls{ppm}, showcasing how utilizing a user privacy policy can facilitate the \gls{put}, as advocated by \name. 

We determine that Swellfish privacy notably benefits the \gls{put} in \name 
being instantiated to protect the private pattern of ``taking medicine'' by Bob over multiple days.
Thus, a \gls{har} application that is curious about Bob's health condition cannot discern if Bob takes medicine on a specific day. 
As in~\cite{tex2022swellfish}, the private pattern is revealed via \textit{count queries} over predefined Bob's events, like $[walk, swallow, \\ drink, lay \ down]$, where each query returns the occurrences of such events, e.g., $[1, 0, 0, 0]$, performed by Bob at a point in time. 

\begin{table}
\scriptsize
\centering
 \caption{Instantiating Swellfish privacy to conceal private patterns in \name.}
        \label{tab:swellfish}
  \vspace{-2ex}
  \begin{tabular}{|c|*{7}{c|}}
  
	\toprule
	\diagbox{Day}{Event}{Time} & \cellcolor{gr!75}{$t_1$}  & \cellcolor{gr!75}{$t_2$}  &  \cellcolor{gr!75}{$t_3$}  & \cellcolor{gr!75}{$t_4$} & $t_5$ & $t_6$ & $t_7$ \\
	\midrule
	Bob's Day 1 & \cellcolor{pp!75}{swallow} & \cellcolor{pp!75}{drink} & \cellcolor{pp!75}{lay d.} & drink & swallow & lay d. & walk   \\
    Bob's Day 2 & walk & \cellcolor{pp!75}{swallow} & \cellcolor{pp!75}{drink} & \cellcolor{pp!75}{lay d.} & walk & lay d. & drink   \\
    ... & ... & ... & ... & ... & ... & ... & ... \\
    Bob's Day n & \cellcolor{pp!75}{swallow} & \cellcolor{pp!75}{drink} & \cellcolor{pp!75}{lay d.} & walk & \cellcolor{pp!75}{swallow} & \cellcolor{pp!75}{drink} & \cellcolor{pp!75}{lay d.} \\
    \hline 
    \noalign{\vskip 0.5mm} 
    $Q(t)$ & -- & -- & 2 & 1 & 0 & 0 & 1 \\ 
    \noalign{\vskip 0.5mm} 
    \hline
    \noalign{\vskip 0.5mm} 
    $Lap(\cdot)$ & $\frac{n\cdot\epsilon}{3}$ & $\frac{n\cdot\epsilon}{3}$ & $\frac{n\cdot\epsilon}{3}$ & $\frac{n\cdot\epsilon}{3}$ & $\frac{n\cdot\epsilon}{2}$ & $n\cdot\epsilon$ & 0 \\
	\bottomrule
\end{tabular}
  \centering
    \center{\scriptsize\colorbox{gr!75}{\parbox{.005\textwidth}{\transparent{0.0}\textcolor{gr!75}{0}}} -- relevance interval; \scriptsize\colorbox{pp!75}{\parbox{.005\textwidth}{\transparent{0.0}\textcolor{pp!75}{0}}} -- private pattern; \underline{small} $Lap(\cdot)$ means \underline{more} noise, n -- num. of days.} 
\vspace{-3.5ex}
\end{table}

\autoref{tab:swellfish} shows how \name hides the private pattern of ``taking medicine'' utilizing the adapted \gls{dp}-based \gls{ppm} of Swellfish privacy.
The main idea is to inject Laplacian noise to the query result $Q(t)$ \textit{only within} the relevance interval inside which the ``taking medicine'' pattern occurs (as a sequence of ``swallow'' $\rightarrow$ ``drink'' $\rightarrow$ ``lay down'' events) to protect it, while  adding no noise to the result outside the relevance interval to preserve utility. 
This allows us to prevent the detection of at least one simple event, like ``drink'', hence concealing the private pattern.  
Specifically, $M(t) = Q(t) + Lap(\cdot)$ is the \textit{sanitized query result} utilizing this \gls{ppm}, where $Lap(\cdot)$ shows the amount of added Laplacian noise.
We set the size of the relevance interval to \textit{four timestamps} to encompass the length of our private pattern (i.e., $w = 3$ in the w-event level \gls{dp} terminology).
Other parameters that include \textit{global query sensitivity} and \textit{privacy budget} $\epsilon$ 
are chosen to be one and the range 0.1--10, respectively, in accord with~\cite{tex2022swellfish}.

The noise reduction from the privacy budget of $\frac{\epsilon}{3}$, equally distributed in the relevance interval $[t_1, t_4]$, to $0$ outside this relevance interval at $t_7$ in $[t_5, t_7]$ improves utility (i.e., the detection accuracy of public complex events, like ``Bob exercises'') without comprising privacy, leading to a better \gls{put}. 
However, the rapid noise reduction outside the relevance interval is not well-studied even in~\cite{tex2022swellfish}---not to mention the \gls{put} behavior in various other \gls{sps}'s use cases.  

\noindent
\textbf{Results: \gls{ac}-based \glspl{ppm}.} We exemplify how \name can achieve \gls{ac} on private patterns. 
For this, it relies on a \textit{query rewriting} component that takes a query submitted by a data consumer (e.g., \gls{har} application) and enforces \textit{privacy action rules} on the query by rewriting it if necessary. 
These privacy action rules define \textit{how} the query is rewritten, and they are instantiated from the user privacy policy with respect to the \gls{ppm} (cf.~\autoref{fig:overview}).
Note that the query rewriting component is the \textit{entry point} of the query in \name, ensuring \textbf{P2} and following best practices for query rewriting in databases~\cite{johnson2020chorus}. 

We show how \name enforces \gls{ac} using our \textit{running example count query}, where ``swallow'' $\rightarrow$ ``drink'' $\rightarrow$ ``lay down'' events must occur, e.g., within a window of two minutes, to reveal the private pattern of ``taking medicine'' by Bob:\footnote{We use Siddhi query language syntax here: \url{https://siddhi.io} [Accessed on 26.05.2023].}

\begin{adjustbox}{width=\linewidth}
\begin{codenv}
define stream TakeMedicineStr (ts long, cnt_swallow int, 
cnt_drink int, cnt_layd int); 
from every e1=TakeMedicineStr[ user_activity == 'swallow' ]
     -> e2=TakeMedicineStr[ user_activity == 'drink' ]
     -> e3=TakeMedicineStr[ user_activity == 'lay down' ]
    within 2 min
select e3.ts, count(e1.user_activity) as cnt_swallow, 
count(e2.user_activity) as cnt_drink, 
count(e3.user_activity) as cnt_layd 
insert into TakeMedicinePattern;
\end{codenv}
\end{adjustbox}

Upon seeing such a query---issued by the \gls{har} application---the query rewriting component of \name first checks if there exist privacy action rules related to this query. 
Imagine Bob has defined a privacy policy rule demanding to forward the results of queries, searching for the ``taking medicine'' pattern, \textit{only} to himself. 
In this case, the \gls{har} application sends the query to detect these patterns, violating Bob's privacy policy.
This violation is noticed by the query rewriting component, triggering the action rule to rewrite the \texttt{@sink} clause such that the query result is only presented to Bob:
\begin{codenv}
@sink(publisher='Bob')
\end{codenv}
With this, Bob's private patterns become protected while preserving utility, i.e., the detection accuracy of his public complex events, like ``Bob exercises'', improving the \gls{put}. 

\noindent
\textbf{Research Opportunity.} 
We identify a few challenges to protection of private patterns in \name. 
First, both \gls{dp}- and \gls{ac}-based \glspl{ppm} need to be benchmarked on the set of private patterns to understand \glspl{ppm}' \textit{efficiency and tradeoffs}, e.g., in terms of protection levels and runtime performance.  
Second, we see potential for these two \glspl{ppm} to \textit{complement each other} to protect private patterns, materializing our \textbf{P3}. 
While we are not aware of any exemplary architectures for combining \glspl{ppm} in \glspl{sps}, there exist some in databases~\cite{johnson2020chorus}. 
Third, we view \textit{generalizability} of private patterns' protection in \name, as another research avenue, like discovering and preserving private patterns in enterprise environments, where such patterns can leak intellectual property. 
This research opportunity is especially topical, given a revived interest in the protection of private patterns~\cite{gu2023differential}.


\vspace{-2ex}
\subsection{Mitigate Invasive Queries} 
\label{subsec:inv-quer}
\noindent
\textbf{\gls{ppm} Selection.} We perceive invasive queries as a complex privacy threat, hence it is unlikely to be addressed by any of the \glspl{ppm} used by \name (i.e., \gls{ml}-, \gls{dp}-, and \gls{ac}-based) in their traditional form alone. 
Combining different \glspl{ppm} as per \textbf{P3} (e.g., \gls{dp} and \gls{ac}) should help to tackle such complex threats, but it remains a difficult task (cf. \autoref{subsec:prv-patt}), with little research done in that direction~\cite{johnson2020chorus}.
Hence, we explore how the \textit{operator placement} in \glspl{sps}, defining \textit{where} and \textit{how} the query operators are processed within the multilayered \gls{sps} \cite{luthra2021}, can be leveraged as the \gls{ac}-based \gls{ppm} against \gls{hbc} \gls{sps} nodes (cf.~\autoref{sec:vision}), 
mitigating invasive queries close towards the source of the threat, satisfying \textbf{P2} and implicitly \textbf{P1}. 

\noindent
\textbf{Results: \gls{ac}-based \glspl{ppm}.}
We sketch how the operator placement can be utilized in \name to address invasive queries. 
To date, the operator placement algorithms focus only on \textit{utility-related metrics}, like latency and throughput, to decide which \gls{sps} nodes will process the queries~\cite{luthra2021, nardelli2019efficient}. 
There exists limited work that considers \textit{privacy} to inform the operator placement, e.g.,~\cite{salvaneschi2019language} combines privacy and utility goals in the placement decisions, but only from the viewpoint of designing a query language for \glspl{sps}, while~\cite{dwarakanath2017trustcep} incorporates user's trust into an \gls{sps} node to the cost model, however assessing trust of resources in a multilayered \gls{sps}, whose properties are hidden from end users, is infeasible.
In contrast, \name approaches the task of operator placement from the point of balancing a \gls{put}. 
Specifically, we seek to reuse existing operator placement algorithms optimizing utility and augment them with privacy by reducing the search space for the operator placement to \textit{trusted nodes} (cf.~\autoref{fig:overview}). 
\name relies on privacy action rules derived from the user's privacy policy applied with respect to the \gls{ppm} to decide on nodes' trustworthiness, like trusted sinks specified by \gls{ac}-based \glspl{ppm} (cf.~\autoref{subsec:prv-patt}) which limit query processing to, e.g., Bob's or his family member's devices. 

\noindent
\textbf{Research Opportunity.}
Whilst we outline the idea of reducing the search space for the operator placement to address invasive queries, the question of \textit{how to} derive information on nodes' trustworthiness from the action rules and encapsulate it in existing placement algorithms remains open. 
Thus, another research avenue is to explore factors for determining trusted nodes in a multilayered \gls{sps}, like (1) support of specific \glspl{ppm} verifiable by \gls{sps}'s users, (2) capability of running \glspl{ppm} inside \glspl{tee}, (3) trust perception of an \gls{sps} node provided by its peers~\cite{salajegheh2022cora}, and (4) levels of transparency, e.g., reliance on open architectures as well as voluntary certification. 

%% file: section/conclusion.tex
\glsresetall

\vspace{-2ex}
\section{Conclusion}
In this work, we introduce our vision for \name, enabling holistic privacy protection in multilayered \glspl{sps} with a focus on balancing a \gls{put}. 
\name utilizes critical privacy threats in multilayered \glspl{sps} that we identify in order to systematically select, customize, and apply state-of-the-art \glspl{ppm} to address these critical threats while promoting the \gls{put}.  
Furthermore, we provide research opportunities for the found privacy threats and \glspl{ppm} tackling them, outlining how such \glspl{ppm} can interact with core \gls{sps}'s mechanisms, like query rewriting and operator placement.

%% file: section/acknowledgement.tex
\vspace{-2ex}
\section{Acknowledgments}
\label{sec:ack}
We are thankful to Majid Lotfian Delouee, He Gu, Rana Tallal Javed, and Espen Volnes for the initial discussions on this paper.
We would like to thank Christine Sch{\"a}ler and Martin Sch{\"a}ler for their input on Swellfish privacy, and the anonymous reviewers for giving valuable feedback. 
This work has been co-funded by the Research Council of Norway as part of the project Parrot (311197), DFG Collaborative Research Center (CRC) 1053–MAKI, and DFKI Darmstadt.

%% file: paper.bbl

\begin{thebibliography}{29}


\ifx \showCODEN    \undefined \def \showCODEN     #1{\unskip}     \fi
\ifx \showDOI      \undefined \def \showDOI       #1{#1}\fi
\ifx \showISBNx    \undefined \def \showISBNx     #1{\unskip}     \fi
\ifx \showISBNxiii \undefined \def \showISBNxiii  #1{\unskip}     \fi
\ifx \showISSN     \undefined \def \showISSN      #1{\unskip}     \fi
\ifx \showLCCN     \undefined \def \showLCCN      #1{\unskip}     \fi
\ifx \shownote     \undefined \def \shownote      #1{#1}          \fi
\ifx \showarticletitle \undefined \def \showarticletitle #1{#1}   \fi
\ifx \showURL      \undefined \def \showURL       {\relax}        \fi
\providecommand\bibfield[2]{#2}
\providecommand\bibinfo[2]{#2}
\providecommand\natexlab[1]{#1}
\providecommand\showeprint[2][]{arXiv:#2}

\bibitem[Biswas and Palamidessi(2022)]%
        {biswas2022three}
\bibfield{author}{\bibinfo{person}{Sayan Biswas} {and}
  \bibinfo{person}{Catuscia Palamidessi}.} \bibinfo{year}{2022}\natexlab{}.
\newblock \showarticletitle{{PRIVIC: A Privacy-preserving Method for
  Incremental Collection of Location Data}}.
\newblock \bibinfo{journal}{\emph{arXiv preprint arXiv:2206.10525}}
  (\bibinfo{year}{2022}).
\newblock


\bibitem[Bloom(2022)]%
        {bloom2022privacy}
\bibfield{author}{\bibinfo{person}{Cara Bloom}.}
  \bibinfo{year}{2022}\natexlab{}.
\newblock \bibinfo{title}{{Privacy Threat Modeling}}.
\newblock
\newblock
\newblock
\shownote{\url{https://www.usenix.org/conference/pepr22/presentation/bloom}}.


\bibitem[Burkhalter et~al\mbox{.}(2020)]%
        {burkhalter2020timecrypt}
\bibfield{author}{\bibinfo{person}{Lukas Burkhalter}, \bibinfo{person}{Anwar
  Hithnawi}, \bibinfo{person}{Alexander Viand}, \bibinfo{person}{Hossein
  Shafagh}, {and} \bibinfo{person}{Sylvia Ratnasamy}.}
  \bibinfo{year}{2020}\natexlab{}.
\newblock \showarticletitle{{TimeCrypt: Encrypted Data Stream Processing at
  Scale with Cryptographic Access Control}}. In
  \bibinfo{booktitle}{\emph{Proceedings of the 17th USENIX Symposium on
  Networked Systems Design and Implementation}}. \bibinfo{pages}{835--850}.
\newblock


\bibitem[Cao et~al\mbox{.}(2010)]%
        {cao2010castle}
\bibfield{author}{\bibinfo{person}{Jianneng Cao}, \bibinfo{person}{Barbara
  Carminati}, \bibinfo{person}{Elena Ferrari}, {and} \bibinfo{person}{Kian-Lee
  Tan}.} \bibinfo{year}{2010}\natexlab{}.
\newblock \showarticletitle{{Castle: Continuously Anonymizing Data Streams}}.
\newblock \bibinfo{journal}{\emph{IEEE Transactions on Dependable and Secure
  Computing}} \bibinfo{volume}{8}, \bibinfo{number}{3} (\bibinfo{year}{2010}),
  \bibinfo{pages}{337--352}.
\newblock


\bibitem[Chen et~al\mbox{.}(2017)]%
        {chen2017pegasus}
\bibfield{author}{\bibinfo{person}{Yan Chen}, \bibinfo{person}{Ashwin
  Machanavajjhala}, \bibinfo{person}{Michael Hay}, {and}
  \bibinfo{person}{Gerome Miklau}.} \bibinfo{year}{2017}\natexlab{}.
\newblock \showarticletitle{{PeGaSus: Data-adaptive Differentially Private
  Stream Processing}}. In \bibinfo{booktitle}{\emph{Proceedings of the 2017 ACM
  SIGSAC Conference on Computer and Communications Security}}.
  \bibinfo{pages}{1375--1388}.
\newblock


\bibitem[Dwarakanath et~al\mbox{.}(2017)]%
        {dwarakanath2017trustcep}
\bibfield{author}{\bibinfo{person}{Rahul Dwarakanath}, \bibinfo{person}{Boris
  Koldehofe}, \bibinfo{person}{Yashas Bharadwaj}, \bibinfo{person}{The An~Binh
  Nguyen}, \bibinfo{person}{David Eyers}, {and} \bibinfo{person}{Ralf
  Steinmetz}.} \bibinfo{year}{2017}\natexlab{}.
\newblock \showarticletitle{{TrustCEP: Adopting a Trust-based Approach for
  Distributed Complex Event Processing}}. In \bibinfo{booktitle}{\emph{2017
  18th IEEE International Conference on Mobile Data Management (MDM)}}.
  \bibinfo{pages}{30--39}.
\newblock


\bibitem[Erdemir et~al\mbox{.}(2022)]%
        {erdemir2022active}
\bibfield{author}{\bibinfo{person}{Ecenaz Erdemir}, \bibinfo{person}{Pier~Luigi
  Dragotti}, {and} \bibinfo{person}{Deniz Gunduz}.}
  \bibinfo{year}{2022}\natexlab{}.
\newblock \showarticletitle{{Active Privacy-Utility Trade-off Against Inference
  in Time-Series Data Sharing}}.
\newblock \bibinfo{journal}{\emph{arXiv preprint arXiv:2202.05833}}
  (\bibinfo{year}{2022}).
\newblock


\bibitem[Fioretto and Van~Hentenryck(2019)]%
        {fioretto2019optstream}
\bibfield{author}{\bibinfo{person}{Ferdinando Fioretto} {and}
  \bibinfo{person}{Pascal Van~Hentenryck}.} \bibinfo{year}{2019}\natexlab{}.
\newblock \showarticletitle{{OptStream: Releasing Time Series Privately}}.
\newblock \bibinfo{journal}{\emph{Journal of Artificial Intelligence Research}}
   \bibinfo{volume}{65} (\bibinfo{year}{2019}), \bibinfo{pages}{423--456}.
\newblock


\bibitem[Gu et~al\mbox{.}(2023)]%
        {gu2023differential}
\bibfield{author}{\bibinfo{person}{He Gu}, \bibinfo{person}{Thomas Plagemann},
  \bibinfo{person}{Maik Benndorf}, \bibinfo{person}{Vera Goebel}, {and}
  \bibinfo{person}{Boris Koldehofe}.} \bibinfo{year}{2023}\natexlab{}.
\newblock \showarticletitle{{Differential Privacy for Protecting Private
  Patterns in Data Streams}}.
\newblock \bibinfo{journal}{\emph{arXiv preprint arXiv:2305.06105}}
  (\bibinfo{year}{2023}).
\newblock


\bibitem[Guo and Zhang(2013)]%
        {guo2013fast}
\bibfield{author}{\bibinfo{person}{Kun Guo} {and} \bibinfo{person}{Qishan
  Zhang}.} \bibinfo{year}{2013}\natexlab{}.
\newblock \showarticletitle{{Fast Clustering-based Anonymization Approaches
  with Time Constraints for Data Streams}}.
\newblock \bibinfo{journal}{\emph{Knowledge-Based Systems}}
  \bibinfo{volume}{46} (\bibinfo{year}{2013}), \bibinfo{pages}{95--108}.
\newblock


\bibitem[Hajihassnai et~al\mbox{.}(2021)]%
        {hajihassnai2021obscurenet}
\bibfield{author}{\bibinfo{person}{Omid Hajihassnai}, \bibinfo{person}{Omid
  Ardakanian}, {and} \bibinfo{person}{Hamzeh Khazaei}.}
  \bibinfo{year}{2021}\natexlab{}.
\newblock \showarticletitle{{ObscureNet: Learning Attribute-invariant Latent
  Representation for Anonymizing Sensor Data}}. In
  \bibinfo{booktitle}{\emph{Proceedings of the International Conference on
  Internet-of-Things Design and Implementation}}. \bibinfo{pages}{40--52}.
\newblock


\bibitem[{IBM Corporation}(2023)]%
        {ibm2023query}
\bibfield{author}{\bibinfo{person}{{IBM Corporation}}.}
  \bibinfo{year}{2023}\natexlab{}.
\newblock \bibinfo{title}{{Query Rewrite}}.
\newblock
\newblock
\newblock
\shownote{\url{https://www.ibm.com/docs/en/guardium/11.2?topic=protect-query-rewrite}}.


\bibitem[Jin et~al\mbox{.}(2022)]%
        {jin2022exploring}
\bibfield{author}{\bibinfo{person}{Haojian Jin}, \bibinfo{person}{Boyuan Guo},
  \bibinfo{person}{Rituparna Roychoudhury}, \bibinfo{person}{Yaxing Yao},
  \bibinfo{person}{Swarun Kumar}, \bibinfo{person}{Yuvraj Agarwal}, {and}
  \bibinfo{person}{Jason~I Hong}.} \bibinfo{year}{2022}\natexlab{}.
\newblock \showarticletitle{{Exploring the Needs of Users for Supporting
  Privacy-Protective Behaviors in Smart Homes}}. In
  \bibinfo{booktitle}{\emph{CHI Conference on Human Factors in Computing
  Systems}}. \bibinfo{pages}{1--19}.
\newblock


\bibitem[Johnson et~al\mbox{.}(2020)]%
        {johnson2020chorus}
\bibfield{author}{\bibinfo{person}{Noah Johnson}, \bibinfo{person}{Joseph~P
  Near}, \bibinfo{person}{Joseph~M Hellerstein}, {and} \bibinfo{person}{Dawn
  Song}.} \bibinfo{year}{2020}\natexlab{}.
\newblock \showarticletitle{{Chorus: A Programming Framework for Building
  Scalable Differential Privacy Mechanisms}}. In \bibinfo{booktitle}{\emph{2020
  IEEE European Symposium on Security and Privacy (EuroS\&P)}}.
  \bibinfo{pages}{535--551}.
\newblock


\bibitem[Luthra et~al\mbox{.}(2021)]%
        {luthra2021}
\bibfield{author}{\bibinfo{person}{Manisha Luthra}, \bibinfo{person}{Boris
  Koldehofe}, \bibinfo{person}{Niels Danger}, \bibinfo{person}{Pascal
  Weisenberger}, \bibinfo{person}{Guido Salvaneschi}, {and}
  \bibinfo{person}{Ioannis Stavrakakis}.} \bibinfo{year}{2021}\natexlab{}.
\newblock \showarticletitle{{TCEP: Transitions in Operator Placement to Adapt
  to Dynamic Network Environments}}.
\newblock \bibinfo{journal}{\emph{J. Comput. System Sci.}}
  \bibinfo{volume}{122} (\bibinfo{year}{2021}), \bibinfo{pages}{94--125}.
\newblock


\bibitem[Nardelli et~al\mbox{.}(2019)]%
        {nardelli2019efficient}
\bibfield{author}{\bibinfo{person}{Matteo Nardelli}, \bibinfo{person}{Valeria
  Cardellini}, \bibinfo{person}{Vincenzo Grassi}, {and}
  \bibinfo{person}{Francesco~Lo Presti}.} \bibinfo{year}{2019}\natexlab{}.
\newblock \showarticletitle{{Efficient Operator Placement for Distributed Data
  Stream Processing Applications}}.
\newblock \bibinfo{journal}{\emph{IEEE Transactions on Parallel and Distributed
  Systems}} \bibinfo{volume}{30}, \bibinfo{number}{8} (\bibinfo{year}{2019}),
  \bibinfo{pages}{1753--1767}.
\newblock


\bibitem[Onica et~al\mbox{.}(2016)]%
        {onica2016confidentiality}
\bibfield{author}{\bibinfo{person}{Emanuel Onica}, \bibinfo{person}{Pascal
  Felber}, \bibinfo{person}{Hugues Mercier}, {and} \bibinfo{person}{Etienne
  Rivi{\`e}re}.} \bibinfo{year}{2016}\natexlab{}.
\newblock \showarticletitle{{Confidentiality-preserving Publish/Subscribe: A
  Survey}}.
\newblock \bibinfo{journal}{\emph{ACM computing surveys (CSUR)}}
  \bibinfo{volume}{49}, \bibinfo{number}{2} (\bibinfo{year}{2016}),
  \bibinfo{pages}{1--43}.
\newblock


\bibitem[Palanisamy et~al\mbox{.}(2018)]%
        {palanisamy2018preserving}
\bibfield{author}{\bibinfo{person}{Saravana~Murthy Palanisamy},
  \bibinfo{person}{Frank D{\"u}rr}, \bibinfo{person}{Muhammad~Adnan Tariq},
  {and} \bibinfo{person}{Kurt Rothermel}.} \bibinfo{year}{2018}\natexlab{}.
\newblock \showarticletitle{{Preserving Privacy and Quality of Service in
  Complex Event Processing through Event Reordering}}. In
  \bibinfo{booktitle}{\emph{Proceedings of the 12th ACM International
  Conference on Distributed and Event-based Systems}}. \bibinfo{pages}{40--51}.
\newblock


\bibitem[Plagemann et~al\mbox{.}(2022)]%
        {plagemann2022towards}
\bibfield{author}{\bibinfo{person}{Thomas Plagemann}, \bibinfo{person}{Vera
  Goebel}, \bibinfo{person}{Matthias Hollick}, {and} \bibinfo{person}{Boris
  Koldehofe}.} \bibinfo{year}{2022}\natexlab{}.
\newblock \showarticletitle{{Towards Privacy Engineering for Real-time
  Analytics in the Human-centered Internet of Things}}.
\newblock \bibinfo{journal}{\emph{arXiv preprint arXiv:2210.16352}}
  (\bibinfo{year}{2022}).
\newblock


\bibitem[Quoc et~al\mbox{.}(2017)]%
        {quoc2017privapprox}
\bibfield{author}{\bibinfo{person}{Do~Le Quoc}, \bibinfo{person}{Martin Beck},
  \bibinfo{person}{Pramod Bhatotia}, \bibinfo{person}{Ruichuan Chen},
  \bibinfo{person}{Christof Fetzer}, {and} \bibinfo{person}{Thorsten Strufe}.}
  \bibinfo{year}{2017}\natexlab{}.
\newblock \showarticletitle{{PrivApprox: Privacy-Preserving Stream Analytics}}.
  In \bibinfo{booktitle}{\emph{Proceedings of the 2017 USENIX Conference on
  Usenix Annual Technical Conference}}. \bibinfo{pages}{659–672}.
\newblock


\bibitem[Ren et~al\mbox{.}(2022)]%
        {ren2022ldpids}
\bibfield{author}{\bibinfo{person}{Xuebin Ren}, \bibinfo{person}{Liang Shi},
  \bibinfo{person}{Weiren Yu}, \bibinfo{person}{Shusen Yang},
  \bibinfo{person}{Cong Zhao}, {and} \bibinfo{person}{Zongben Xu}.}
  \bibinfo{year}{2022}\natexlab{}.
\newblock \showarticletitle{{LDP-IDS: Local Differential Privacy for Infinite
  Data Streams}}. In \bibinfo{booktitle}{\emph{Proceedings of the 2022
  International Conference on Management of Data}}.
  \bibinfo{pages}{1064–1077}.
\newblock


\bibitem[Salajegheh et~al\mbox{.}(2022)]%
        {salajegheh2022cora}
\bibfield{author}{\bibinfo{person}{Mastooreh Salajegheh},
  \bibinfo{person}{Shashank Agrawal}, \bibinfo{person}{Maliheh Shirvanian},
  \bibinfo{person}{Mihai Christodorescu}, {and} \bibinfo{person}{Payman
  Mohassel}.} \bibinfo{year}{2022}\natexlab{}.
\newblock \showarticletitle{{CoRA: Collaborative Risk-aware Authentication}}.
\newblock \bibinfo{journal}{\emph{Cryptology ePrint Archive}}
  (\bibinfo{year}{2022}).
\newblock


\bibitem[Salvaneschi et~al\mbox{.}(2019)]%
        {salvaneschi2019language}
\bibfield{author}{\bibinfo{person}{Guido Salvaneschi}, \bibinfo{person}{Mirko
  K{\"o}hler}, \bibinfo{person}{Daniel Sokolowski}, \bibinfo{person}{Philipp
  Haller}, \bibinfo{person}{Sebastian Erdweg}, {and} \bibinfo{person}{Mira
  Mezini}.} \bibinfo{year}{2019}\natexlab{}.
\newblock \showarticletitle{{Language-integrated Privacy-aware Distributed
  Queries}}.
\newblock \bibinfo{journal}{\emph{Proc. ACM Program. Lang.}}
  \bibinfo{volume}{3}, \bibinfo{number}{OOPSLA} (\bibinfo{year}{2019}),
  \bibinfo{pages}{167--1}.
\newblock


\bibitem[Scopelliti et~al\mbox{.}(2022)]%
        {scopelliti2023end}
\bibfield{author}{\bibinfo{person}{Gianluca Scopelliti},
  \bibinfo{person}{Sepideh Pouyanrad}, \bibinfo{person}{Job Noorman},
  \bibinfo{person}{Fritz Alder}, \bibinfo{person}{Christoph Baumann},
  \bibinfo{person}{Frank Piessens}, {and} \bibinfo{person}{Jan~Tobias
  M\"{u}hlberg}.} \bibinfo{year}{2022}\natexlab{}.
\newblock \showarticletitle{{End-to-End Security for Distributed Event-Driven
  Enclave Applications on Heterogeneous TEEs}}.
\newblock \bibinfo{journal}{\emph{arXiv preprint arXiv:2206.01041}}
  (\bibinfo{year}{2022}).
\newblock


\bibitem[Stach et~al\mbox{.}(2020)]%
        {stach2020bringing}
\bibfield{author}{\bibinfo{person}{Christoph Stach},
  \bibinfo{person}{Cl{\'e}mentine Gritti}, {and} \bibinfo{person}{Bernhard
  Mitschang}.} \bibinfo{year}{2020}\natexlab{}.
\newblock \showarticletitle{{Bringing Privacy Control Back to Citizens:
  DISPEL---A Distributed Privacy Management Platform for the Internet of
  Things}}. In \bibinfo{booktitle}{\emph{Proceedings of the 35th Annual ACM
  Symposium on Applied Computing}}. \bibinfo{pages}{1272--1279}.
\newblock


\bibitem[Sztyler and Stuckenschmidt(2016)]%
        {sztyler2016body}
\bibfield{author}{\bibinfo{person}{Timo Sztyler} {and} \bibinfo{person}{Heiner
  Stuckenschmidt}.} \bibinfo{year}{2016}\natexlab{}.
\newblock \showarticletitle{{On-body Localization of Wearable Devices: An
  Investigation of Position-aware Activity Recognition}}. In
  \bibinfo{booktitle}{\emph{2016 IEEE International Conference on Pervasive
  Computing and Communications (PerCom)}}. \bibinfo{pages}{1--9}.
\newblock


\bibitem[Tex et~al\mbox{.}(2022)]%
        {tex2022swellfish}
\bibfield{author}{\bibinfo{person}{Christine Tex}, \bibinfo{person}{Martin
  Sch{\"a}ler}, {and} \bibinfo{person}{Klemens B{\"o}hm}.}
  \bibinfo{year}{2022}\natexlab{}.
\newblock \showarticletitle{{Swellfish Privacy: Supporting Time-dependent
  Relevance for Continuous Differential Privacy}}.
\newblock \bibinfo{journal}{\emph{Information Systems}} (\bibinfo{year}{2022}),
  \bibinfo{pages}{102079}.
\newblock


\bibitem[Yang and Ardakanian(2022)]%
        {yang2022blinder}
\bibfield{author}{\bibinfo{person}{Xin Yang} {and} \bibinfo{person}{Omid
  Ardakanian}.} \bibinfo{year}{2022}\natexlab{}.
\newblock \showarticletitle{{Blinder: End-to-end Privacy Protection in Sensing
  Systems via Personalized Federated Learning}}.
\newblock \bibinfo{journal}{\emph{arXiv preprint arXiv:2209.12046}}
  (\bibinfo{year}{2022}).
\newblock


\bibitem[Zeuch et~al\mbox{.}(2020)]%
        {zeuch2020nebulastream}
\bibfield{author}{\bibinfo{person}{Steffen Zeuch}, \bibinfo{person}{Ankit
  Chaudhary}, \bibinfo{person}{Bonaventura Del~Monte},
  \bibinfo{person}{Haralampos Gavriilidis}, \bibinfo{person}{Dimitrios
  Giouroukis}, \bibinfo{person}{Philipp~M Grulich}, \bibinfo{person}{Sebastian
  Bre{\ss}}, \bibinfo{person}{Jonas Traub}, {and} \bibinfo{person}{Volker
  Markl}.} \bibinfo{year}{2020}\natexlab{}.
\newblock \showarticletitle{{The NebulaStream Platform for Data and Application
  Management in the Internet of Things}}. In \bibinfo{booktitle}{\emph{10th
  Conference on Innovative Data Systems Research, {CIDR} 2020}}.
\newblock


\end{thebibliography}
